\documentclass[final]{svjour2}
\usepackage{graphicx}
\usepackage{rotating}
\usepackage{amssymb}
\usepackage{mathptmx}
\usepackage[numbers]{natbib}
\makeatletter
\journalname{Journal of Low Temperature Physics}

\bibpunct{}{}{,}{s}{}{,}

\begin{document}

\newcommand{\hdblarrow}{H\makebox[0.9ex][l]{$\downdownarrows$}-}
\title{BCC vs. HCP - The Effect of Crystal Symmetry on the High Temperature Mobility of Solid $^4$He}

\author{A. Eyal         \and
        E. Polturak}

\institute{A. Eyal \and E. Polturak \at
              Department of Physics, Technion, Haifa 32000, Israel \\
              Tel.: +972-4-8292027\\
              Fax: +972-4-8292027\\
              \email{satanan@tx.technion.ac.il}             \\
           \and
           E. Polturak \at
              \email{emilp@physics.technion.ac.il}}

\date{07.02.2012}

\maketitle


\begin{abstract}

We report results of torsional oscillator (TO) experiments on
solid $^4$He at temperatures above 1K. We have previously found
that single crystals, once disordered, show some mobility
(decoupled mass) even at these rather high temperatures. The
decoupled mass fraction with single crystals is typically 20-
30\%. In the present work we performed similar measurements on
polycrystalline solid samples. The decoupled mass with
polycrystals is much smaller, $\sim$ 1\%, similar to what is
observed by other groups. In particular, we compared the
properties of samples grown with the TO's rotation axis at
different orientations with respect to gravity. We found that the
decoupled mass fraction of bcc samples is independent of the angle
between the rotation axis and gravity. In contrast, hcp samples
showed a significant difference in the fraction of decoupled mass
as the angle between the rotation axis and gravity was varied
between zero and 85 degrees. Dislocation dynamics in the solid
offers one possible explanation of this anisotropy.

\end{abstract}

    \section{Introduction} \label{sec:intro}
The physical mechanism responsible for the apparent mobility of
solid He remains a subject of intense study. Experiments done on
solid $^4$He contained inside a torsional oscillator (TO) show a
partial mass decoupling
\cite{KC2004,Reppy2007,Kojima,Davis,Kubota,Shirahama,Golov} below some
200 mK, an indication of some kind of mobility of the solid. The
mass decoupling fraction seen in the various TO experiments is
typically 0.01\%-2\%, except for the unique results of 20\% by
Rittner and Reppy \cite{Reppy2007}. The interpretation of these results is
under debate in terms of several competing models. These include
supersolidity \cite{Anderson,Andreev}, superglass \cite{Balatsky,Andreev_new,Korshunov}, and dislocation mediated effects
\cite{Iwasa,Kuklov,Aleinikava,Reppy2010}.
The possibility of dislocation dynamics came up following measurements of the shear
modulus which showed changes at the same temperature where mass
decoupling was observed \cite{Beamish,Choi}.

Our previous contribution to this subject came through TO
measurements on solid He at higher temperatures, between 1.1K and
1.9K. In contrast to polycrystalline samples used by others, we
grew single crystals inside the sample space of the TO. Crystals
of commercially pure $^4$He or of 100 ppm $^3$He-$^4$He mixtures
were grown at a constant temperature and pressure. We found that
generation of structural disorder within a single crystal caused a
large fraction of the mass to decouple from the TO
\cite{Eyal2010,Eyal2011}. The decoupled mass fraction did not
depend strongly on temperature. The very fact that mass decoupling
in a TO experiment can be observed at practically any temperature
above 1K suggests that the phenomenon is not an usual phase
transition.

One way to distinguish between the  physical scenarios mentioned
above is to look for some signs of anisotropy. The mass decoupling
effect is seen in the response of the solid to stress applied by
the moving wall of the cell. Supersolidity or glassy models of the
solid do not predict any anisotropy of this response. In contrast,
if the response is due to dislocations, one should expect some
anisotropy. For example, at small values of externally applied
stress, it is reasonable to assume that dislocations glide. Glide
occurs on selected crystalline planes, and is naturally sensitive
to the direction of stress relative to the crystalline axes. In
order to check whether this is the case, one needs to vary the
orientation of the crystal with respect to that of the stress.
Since the stress is applied by the walls of the TO, one should
grow crystals at different orientations relative to the rotation
axis of the TO. The experiment described here was designed to test
this scenario.


    \section{Experimental System} \label{sec:experimentalSystem}
\begin{figure}
\begin{center}
\includegraphics[%
  width=1\linewidth,
  keepaspectratio]{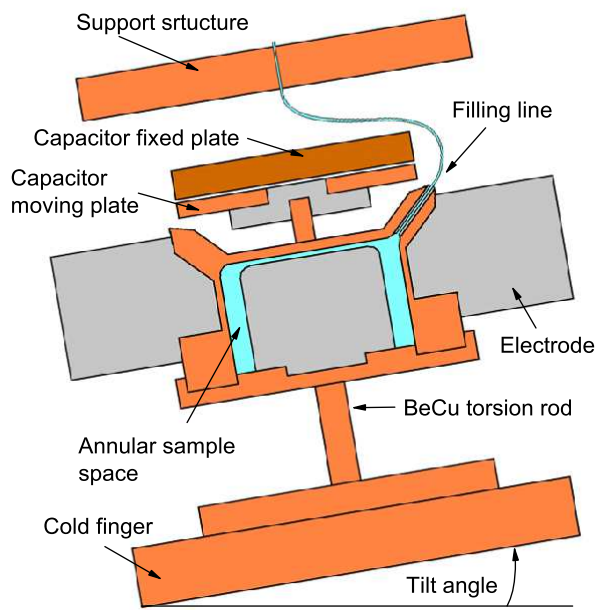}
\end{center}
\caption{Schematic cross section of the torsional
oscillator (TO) assembly. The drawing is to-scale. The assembly
can be tilted between zero and 90 degrees, in order to grow
crystals at different angles with regards to gravity. Note that
the filling line enters the cell at a corner so that the entrance
is always at the highest point of the cell. This prevents the
filling line from becoming blocked with solid before the cell is
full, irrespective of the tilt angle of the TO. The bottom of the
cell is thermally connected to a $^3$He refrigerator. The crystals
are grown in the annular space of outer diameter of 18mm.}
\label{fig:TO_cell}
\end{figure}

The interaction of He atoms with practically any substrate is much
stronger than the He-He interaction. This has important
consequences on the growth of He crystals. The strong interaction
with the substrate causes the typical substrate (inner walls of
the cell in our case) to be coated by a dense, close packed layer
of solid He. Consequently, He crystals always grow on a substrate
consisting of amorphous He and are not sensitive to the material
from which the cell is constructed. The orientation of the crystal
is determined by gravity and by the thermal profile inside the
cell. As a result, Helium crystals tend to grow more or less with
the same crystalline orientation in relation to gravity. For
example, in X-ray studies done by Greywall \cite{Greywall-xray},
it was found that 80\% of 99 hcp crystals grown had their c axis
at an angle of less than 30\% from the horizontal direction.
Similar behavior was observed in neutron scattering experiments
\cite{Bossy-un,Tuvy}, with the c axis of hcp crystals again being
close to horizontal. Regarding bcc crystals, we found that these
typically grow with the [111] direction close to vertical
\cite{Bossy-un,Oshri,Tuvy}. Therefore, experiments on different
crystals in the same cell are usually quite reproducible, as most
of these crystals would grow with the same crystallographic
direction with respect to the cell. In order to measure the
decoupled mass fraction of crystals grown with different
crystalline directions, one has to change the direction of the
stress relative to the crystalline axes. Since the orientation of
the crystals tends to remain fixed in space, one has to change the
orientation of the torsional oscillator (TO). To that end, we
built a TO cell which could be tilted, so that the rotation axis
was not parallel to gravity. We grew the crystals by raising the
pressure in the cell, using an open and slightly heated filling
line. One additional requirement was that the filling line must
enter the cell at the highest point, so that it will not become
blocked with solid before the cell is full. For that, we designed
the cell with the filling line entering it from the top corner.
Figure \ref{fig:TO_cell} shows an in-scale schematic cross section
of the cell. We grew bcc and hcp solids in this cell and tested
the mass decoupling with the cell's rotation axis aligned with the
direction of gravity, and with the rotation axis forming an angle
of 2, 5, and 85 degrees with the direction of gravity.

The cell itself was made of beryllium copper with a Stycast bob in
its center, which defines an annular sample space of 11 mm height
and 2 mm width. The internal volume was 1.2 cm$^3$. Solid He was grown
inside this annular space. All the internal corners were rounded,
with sharp corners remaining only at the bottom of the cell. We used a
capacitive pressure gauge to measure the pressure in situ. The gauge,
seen in figure \ref{fig:TO_cell}, includes one (moving)
capacitor plate attached to the top of the cell and another (fixed) plate
connected to an external support structure.

In this study, we grew a total of 21 crystals at temperatures
between 1.39K and 1.75K. Most of the samples were grown using
commercially pure $^4$He, but we also grew some samples containing
100ppm of $^3$He. Initially, the cell's resonant period was 2461
$\mu$sec. In order to check whether any of the results depend on
the frequency at which the oscillator operates, we changed the
resonant period of the cell from 2461 $\mu$sec to 4388 $\mu$sec by
reducing the diameter of the torsion rod. In order to eliminate
any possible influence of the oscillatory motion of the TO on
crystal growth, several crystals were grown with the TO turned
off. All these samples showed the same mass decoupling fraction as
the samples grown with the TO oscillating.


\section{Results} \label{sec:results}

\begin{figure}
\begin{center}
\includegraphics[%
  width=1.2\linewidth,
  keepaspectratio]{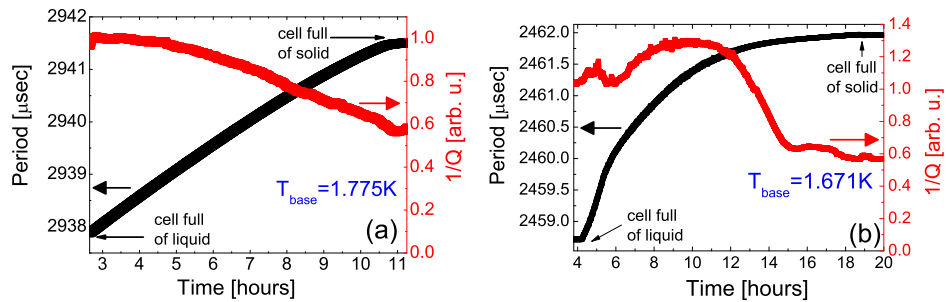}
\end{center}
\caption{Time dependence of the period and
dissipation of the TO during solid growth. The black symbols in
both plots are the resonant period of the TO, which at the
beginning is that of a cell with no solid, and at the end of the
growth process is that of a cell full of solid helium. The red
symbols show the dissipation during this process. Panel (a) shows
the period and dissipation during growth of a single crystal in a
TO aligned with gravity which has a filling line entering it from
the top. In this case, the solid grows as a single crystal. Panel
(b) shows the period and dissipation of a TO tilted at an angle of
85 degrees with respect to the direction of gravity. The time
dependence of the period in (b) is consistent with a
polycrystalline growth.} \label{fig:85deg_growth}
\end{figure}

The solid He samples grown in the TO cell were all
polycrystalline. The likely reason for that is the temperature
profile inside the cell, which was not symmetric about the
rotation axis of the TO. The hottest point in the cell was the
entry point of the heated filling line, which was off-center for
reasons explained above.  For comparison, in panel (a) of figure
\ref{fig:85deg_growth} we show the time dependence of the period
and dissipation of the TO during growth of a single crystal. These
data were taken with the cell used in our previous work
\cite{Eyal2011} in which the rotation axis was parallel to
gravity, and the filling line entered the cell at the top. The
approximately linear increase of the period with time is
consistent with the crystal growing from the bottom of the cell,
and gradually filling the sample space with the liquid-solid
interface horizontal. During the growth, the dissipation of the TO
decreases continuously, indicating that the crystal is of good
quality with low internal friction. Turning to panel (b) of figure
\ref{fig:85deg_growth}, here we show similar data for a TO rotated
at 85 degrees. These data were taken during the current
experiment. The dependence of the period on time is consistent
with the solid growing around the circumference of the sample
space, filling the cell from the outside towards the center. At
the beginning of the growth, the dissipation of the TO increases.
This indicates that many solid grains are created, and there is
internal friction between them. Obviously, this type of growth
results in a polycrystal. The resonant period change upon growth
in the tilted cell was 3.25 $\mu$ sec, about the change expected
from the classical moment of inertia. For completeness, we mention
that the solid in both experiments shown in figure \ref{fig:85deg_growth}
was grown by applying a small constant over-pressure.

At the end of the growth, the entrance to the cell becomes blocked
with solid and the temperature profile changes abruptly. This
causes stress, and leads to additional structural disorder in the
sample. As before \cite{Eyal2010}, we observed that some fraction
of the mass decoupled from the TO. All the polycrystalline samples
showed some mass decoupling, the fraction ranging between 0.1\%
and 1.3\%. The initial mass decoupling fraction depended on the
intensity of the pressure/temperature step which disordered the
solid. After cooling the TO, the mass decoupling fraction reached
a limiting value for each direction and crystal symmetry.

For the bcc phase, this limiting value was around 1.3\%,
independent of the orientation of the TO. For the hcp phase, there
were differences between samples grown with different orientations
of the TO. Hcp samples grown with the rotation axis aligned with
gravity, or tilted at an angle of at most 5 degrees, reached the
same mass decoupling fraction as the bcc crystals. Hcp crystals
grown in the TO tilted at an angle 85 degrees reached a limiting
value of 0.75\% - about half of the value of the bcc phase. These
values are the typical results obtained with 21 different crystals,
15 of them grown
as bcc (2 grown at 0 degrees tilt, 3 at 2 degrees, 5 at 5 degrees,
and 5 at 85 degrees), and 6 grown as hcp (3 at 5 degrees and 3 at
85 degrees tilt). There was no apparent difference between the 0,
2 and 5 degree tilts, and only the 85 degree tilt gave a
significant difference between the hcp and bcc crystals.

We found that the size of the mass
decoupling was the same for crystals grown using commercially pure
$^4$He and for those grown using a mixture containing 100ppm of
$^3$He. Similarly, we found that the results did not depend on the
period of the TO.

Comparing these mass decoupling values to those obtained in our
previous experiments \cite{Eyal2010,Eyal2011} using single
crystals, we see a huge difference. Except for the solid being
polycrystalline in the current experiment and a single crystal in
our previous work, we used the same cryostat, materials and growth
methods in both experiments. Therefore, it seems that
polycrystallinity is the cause of the difference in decoupled
mass. We mention that the decoupled mass fraction with
polycrystals is very similar to the one seen in low temperature
experiments using $^4$He of commercial
purity \cite{KC2004,Reppy2007,Kojima,Davis,Kubota,Shirahama,Golov}.

\begin{figure}
\begin{center}
\includegraphics[%
  width=1\linewidth,
  keepaspectratio]{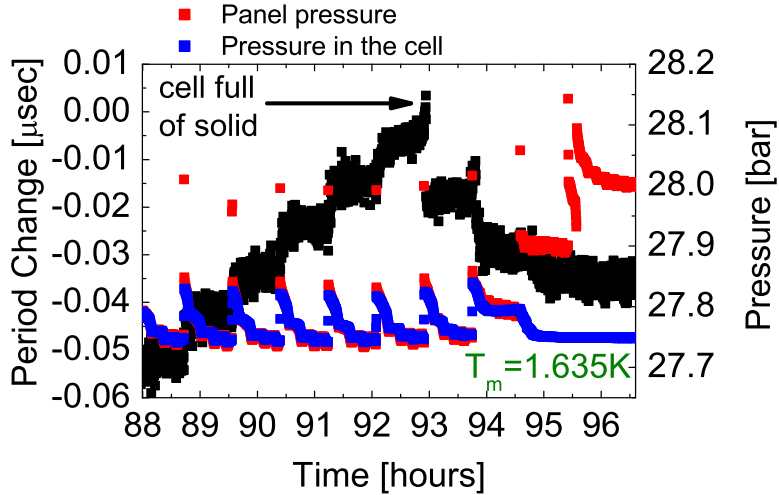}
\end{center}
\caption{Mass decoupling of a solid sample grown as
bcc. In this experiment the TO's rotation axis was tilted by 85 degrees
with respect to gravity. The figure shows the final stages of the
growth process, followed by mass decoupling (decrease of the
period). The decoupled mass fraction reached a constant value
 of 1\% after several hours of relaxation. Red and blue
symbols show the externally applied pressure and the pressure
inside the cell respectively.  }
\label{fig:tilted_break_bcc}
\end{figure}

\begin{figure}
\begin{center}
\includegraphics[%
  width=0.9\linewidth,
  keepaspectratio]{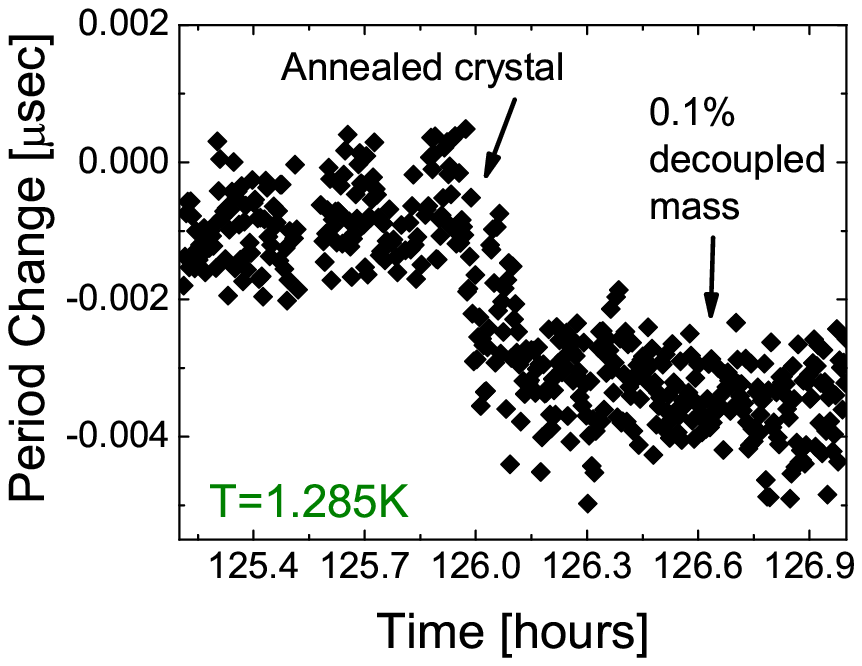}
\end{center}
\caption{Spontaneous mass decoupling of an hcp
solid sample. This sample was initially grown as bcc and then
cooled through the triple point to the hcp phase. The TO's
rotation axis was tilted by 85 degrees. Upon further cooling of
the solid shown in figure \ref{fig:tilted_break_bcc}, as it
reached the lower triple point, it annealed and the decoupled mass
fraction became zero. The crystal followed the melting line down
to a temperature of around 1.285K, where the cell became full of
solid, and spontaneous mass decoupling took place. In this case,
the decoupled mass fraction reached a much lower value, of 0.1\%.
Note that the vertical scale in this figure is about a factor of
10 finer than that in figure \ref{fig:tilted_break_bcc}. This
illustrates the different behavior of bcc and hcp solids in the
tilted TO. } \label{fig:tilted_break_hcp}
\end{figure}

To reduce the influence of the initial disordering process, we
grew bcc samples and then cooled them through the bcc-hcp
transition into the hcp phase. Figures \ref{fig:tilted_break_bcc} and \ref{fig:tilted_break_hcp} show
such data. In this case the solid was grown by periodically adding
small amounts of He to the cell. Figure \ref{fig:tilted_break_bcc}
shows the final stages
of the growth of a bcc solid. Once the cell is full (at 93 hours),
mass decoupling takes place and the period of the TO decreases.
This mass decoupling was produced by applying several pressure
pulses to a cell filled with solid. The first pressure pulse after
the cell is completely full of solid only begins solidifying the
helium in the filling line, so the pressure is still transmitted
into the cell. The following pressure pulses are with the filling
line progressively blocked with solid. These pulses cannot change
the pressure inside the cell directly, but can still cause stress
due to change of the heat flux into the cell, changing the
temperature profile within. At this stage, the sample gradually
cools to the set temperature of the TO stage. This is the reason
why the period of the TO continues to decrease even after the
filling line is blocked. This is evident from the pressure in
the cell. As can be seen in the figure, after 94.5 hours the filling
line is blocked, and subsequent external pressure pulses no longer
influence the pressure inside the cell.

\begin{figure}
\begin{center}
\includegraphics[%
  width=0.75\linewidth,
  keepaspectratio]{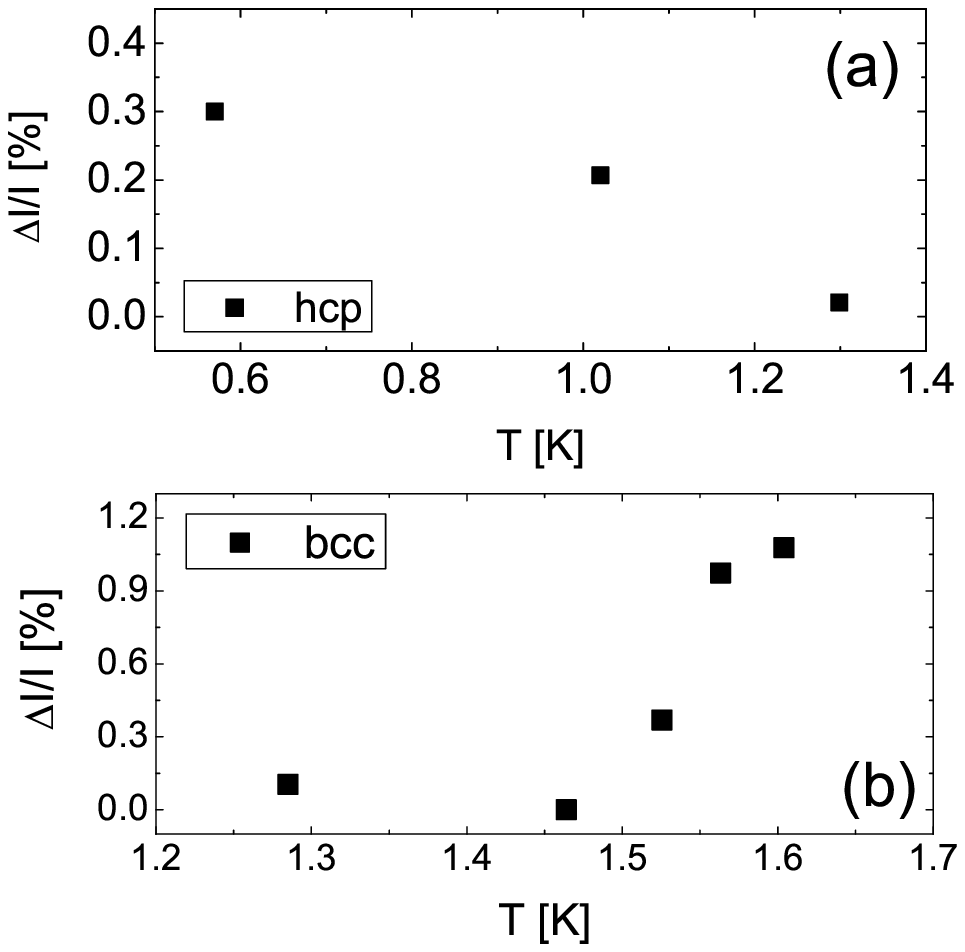}
\end{center}
\caption{Temperature dependence of the mass
decoupling for two samples grown in a TO with a rotation axis
tilted by 85 degrees with respect to gravity. Panel (a)  an hcp
solid grown at 1.419K, and panel (b)  a bcc solid grown at 1.635K,
cooled through the bcc-hcp coexistence region down into the hcp
phase.} \label{fig:temperature_dependence}
\end{figure}

Panel (b) of figure \ref{fig:temperature_dependence} shows the what happens
once this sample is cooled. The bcc solid gradually converts into
the hcp phase during cooldown. It is seen that the decoupled mass
fraction decreases during this stage, and vanishes at the lower
triple point of 1.46K. We remark that bcc crystals grown at
temperatures less than 1.7K, inevitably pass through the lower
triple point during cooldown. At the triple point, the bcc to hcp
conversion is done in the presence of liquid. This transformation
occurs spontaneously inside an isolated cell (the filling line is
blocked with solid). If there is any mass decoupling associated
with this process, it is independent of external factors.

In figure \ref{fig:tilted_break_hcp} we show the spontaneous
disordering of the hcp solid created this way, after the cell was
cooled further to 1.285K. As the sample cooled, the fluid in the
cell gradually solidified. At 1.285K, the cell was full of
hcp solid which was then disordered due to thermal stress and
showed mass decoupling. This sequence of events took place
irrespective of the orientation of the TO. We can therefore
compare the size of the mass decoupling in the bcc and hcp phases,
for crystals that underwent the same procedure, but were grown
with the TO at different orientations. When the TO's rotation axis
was aligned with the direction of gravity, or at most 5 degrees
from it, the bcc phase and the hcp phase showed the same maximal
mass decoupling value within experimental error. On the other
hand, when the rotation axis was at an angle of 85 degrees from
the direction of gravity, the bcc solid reached a maximal mass
decoupling value of around 1\%, whereas the hcp reached only
0.1\%, as shown in figure \ref{fig:tilted_break_hcp}.

Figure \ref{fig:temperature_dependence} shows the temperature
dependence of the mass decoupling of two different crystals grown
in a cell tilted at an angle of 85 degrees. As can be seen, the
hcp crystal reaches much lower values than the bcc one, even after
significant cooling. The results for the 0-5 degree tilt give the
same temperature dependence reported in our previous work
\cite{Eyal2011}, except for the values of the decoupled mass being
much smaller.

\section{Discussion} \label{sec:discussion}

The strong dependence of the decoupled mass fraction on the growth
direction in the anisotropic hcp crystal might be an indication
that dislocations play an important role in the phenomenon. Bcc
crystals are isotropic and there are many (\{110\}, \{112\}, and
\{123\}) equivalent slip planes. If the mass decoupling effect is
associated with dislocation glide, we expect that in the bcc solid
the coupling to the TO will show little dependence on the angle
between the rotation axis and the crystallographic direction. In
the hcp crystal, on the other hand, there is only one easy slip
plane for edge dislocations, (0001). Therefore, dislocations would
glide easily in the bcc structure no matter what the direction of
stress is, while in the hcp structure the glide will be much more
sensitive to the direction of stress in space, which is determined
by the orientation of the cell.

The small amount of decoupled mass seen in our experiment using
polycrystalline samples is similar to that observed in the low
temperature TO experiments on solid helium
\cite{KCScience,Choi,Reppy2007,Kojima,Davis,Kubota,Shirahama}.
In the low temperature experiments the crystals were
grown using a blocked capillary method which results in
polycrystalline samples \cite{Balibar-disorder}, with the
exception of one experiment \cite{Clark}, in which the cell
contained sharp corners, which again makes it impossible to fill
it with a single crystal.

To conclude, we found that polycrystals grown as bcc always gave
the same mass decoupling fraction, regardless of the growth
direction. Hcp polycrystals, on the other hand, showed much
smaller mass decoupling values when grown perpendicular to the
direction of the rotation axis. Dislocation glide could be
responsible for the apparent anisotropy associated with the mass
decoupling inside the TO.


\begin{acknowledgements}
Technical assistance by S. Hoida, L. Yumin, and A. Post is gratefully acknowledged. This work was supported by the Israel Science Foundation and by the Technion Fund for Research.
\end{acknowledgements}

\pagebreak

\bibliographystyle{spphys}       
\bibliography{supersolid}   

\begin{thebibliography}{10}
\providecommand{\url}[1]{{#1}}
\providecommand{\urlprefix}{URL }
\expandafter\ifx\csname urlstyle\endcsname\relax
  \providecommand{\doi}[1]{DOI \discretionary{}{}{}#1}\else
  \providecommand{\doi}{DOI \discretionary{}{}{}\begingroup
  \urlstyle{rm}\Url}\fi

\bibitem{KC2004}
E.~Kim, M.H.W. Chan, Nature \textbf{427}, 225 (2004)

\bibitem{Reppy2007}
A.S.C. Rittner, J.D. Reppy, Phys. Rev. Lett. \textbf{98}(17), 175302 (2007)

\bibitem{Kojima}
Y.~Aoki, J.C. Graves, H.~Kojima, Phys. Rev. Lett. \textbf{99}(1), 015301 (2007)

\bibitem{Davis}
B.~Hunt, E.~Pratt, V.~Gadagkar, M.~Yamashita, A.V. Balatsky, J.C. Davis,
  Science \textbf{324}, 632 (2009)

\bibitem{Kubota}
A.~Penzev, Y.~Yasuta, M.~Kubota, J. of Low Temp. Phys. \textbf{148}, 677 (2007)

\bibitem{Shirahama}
M.~Kondo, S.~Takada, Y.~Shibayama, K.~Shirahama, J. of Low Temp. Phys.
  \textbf{148}, 695 (2007)

\bibitem{Golov}
D.E. Zmeev, A.I. Golov, arXiv:1104.4555v2  (2011)

\bibitem{Anderson}
P.W. Anderson, Science \textbf{324}, 631 (2009)

\bibitem{Andreev}
A.~Andreev, I.~Lifshitz, Zh. Eksp. Teor. Fiz. \textbf{56}, 2057 (1969)

\bibitem{Balatsky}
Z.~Nussinov, A.V. Balatsky, M.J. Graf, S.A. Trugman, Phys. Rev. B
  \textbf{76}(1), 014530 (2007)

\bibitem{Andreev_new}
A.~Andreev, J. of Exp. and Theo. Phys. \textbf{109}, 103 (2009)

\bibitem{Korshunov}
S.E. Korshunov, JETP Lett. \textbf{90}, 156 (2009)

\bibitem{Iwasa}
I.~Iwasa, Phys. Rev. B \textbf{81}, 104527 (2010)

\bibitem{Kuklov}
L.~Pollet, M.~Boninsegni, A.B. Kuklov, N.V. Prokof'ev, B.V. Svistunov,
  M.~Troyer, Phys. Rev. Lett. \textbf{101}(9), 097202 (2008)

\bibitem{Aleinikava}
D.~Aleinikava, A.B. Kuklov, Phys. Rev. Lett. \textbf{106}, 235302 (2011)

\bibitem{Reppy2010}
J.D. Reppy, Phys. Rev. Lett. \textbf{104}(25), 255301 (2010)

\bibitem{Beamish}
J.~Day, J.~Beamish, Nature \textbf{450}, 853 (2007)

\bibitem{Choi}
H.~Choi, D.~Takahashi, K.~Kono, E.~Kim, Science \textbf{330}, 1512 (2010)

\bibitem{Eyal2010}
A.~Eyal, O.~Pelleg, L.~Embon, E.~Polturak, Phys. Rev. Lett. \textbf{105}(2),
  025301 (2010)

\bibitem{Eyal2011}
A.~Eyal, E.~Polturak, J. Low Temp. Phys. \textbf{163}, 262 (2011)

\bibitem{Greywall-xray}
D.S. Greywall, Phys. Rev. A \textbf{3}, 2106 (1971)

\bibitem{Bossy-un}
J.~Bossy, unpublished

\bibitem{Tuvy}
T.~Markovitz, E.~Polturak, J. Low Temp. Phys. \textbf{123}, 53 (2001)

\bibitem{Oshri}
O.~Pelleg, M.~Shay, S.G. Lipson, E.~Polturak, J.~Bossy, J.C. Marmeggi,
  K.~Horibe, E.~Farhi, A.~Stunault, Phys. Rev. B \textbf{73}(2), 024301 (2006)

\bibitem{KCScience}
E.~Kim, M.H.W. Chan, Science \textbf{305}, 1941 (2004)

\bibitem{Balibar-disorder}
S.~Sasaki, F.~Caupin, S.~Balibar, J. Low Temp. Phys. \textbf{153}, 43 (2008)

\bibitem{Clark}
A.C. Clark, J.T. West, M.H.W. Chan, Phys. Rev. Lett. \textbf{99}, 135302 (2007)

\end{thebibliography}




\end{document}